\input{aipcheck}

\documentclass[
    ,final            
  ]
  {aipproc}

\layoutstyle{6x9}

\newcommand\simlt{\lower.5ex\hbox{$\; \buildrel < \over \sim \;$}}
\newcommand\simgt{\lower.5ex\hbox{$\; \buildrel > \over \sim \;$}}

\begin{document}

\title{Neutrino-Driven Mass Loading of GRMHD Outflows}


\author{Amir Levinson}{
  address={School of Physics \& Astronomy, Tel Aviv University,
Tel Aviv 69978, Israel  Levinson@wise.tau.ac.i}
}

\begin{abstract}
A GRMHD model of disk outflows with neutrino-driven mass ejection is presented.
The model is used to calculate the structure of the outflow in the sub-slow magnetosonic 
region and the mass loading of the outflow, under conditions anticipated in 
the central engines of gamma-ray bursts.  
It is concluded that magnetic launching of ultra-relativistic polar outflows 
is in principle possible along low inclination field lines (with respect to the symmetry axis), 
provided the neutrino luminosity is sufficiently low, $L_\nu\simlt10^{52}$ erg s$^{-1}$.
\end{abstract}

\maketitle


\section{Introduction}
Neutrino-driven mass loss plays an important role in the late phases of proto-neutron
star evolution \cite{Duncan86}-\cite{Metzger06}, and in the outflows expelled 
from the hot, dense disks 
in hyper-accreting black hole systems (e.g., Refs.~\cite{Pruet04,Levinson06}).
In weakly magnetized systems those neutrino-driven outflows are subrelativistic,
because the entropy per baryon generated by neutrino absorption at the base of the 
wind ($s/k_B\simlt100$) does not reach the values required to accelerate the outflow to 
high Lorentz factors.  However, in highly magnetized systems the
rotational energy of the system can be extracted magnetically and the 
outflow can in principle accelerate to high Lorentz factors provided the mass loading of
magnetic field lines is sufficiently low \cite{Levinson93,Li92,Contopoulos94,Vlahakis03,McKinney06}.

Magnetic extraction of the rotational energy of accreted matter in collapsars or, alternatively, 
a highly magnetized proto-neutron star is a plausible 
production mechanism for the $\gamma$-ray emitting jets inferred in long GRBs.
To account for the characteristic luminosities and Lorentz factors observed in those
systems two conditions must be fulfilled: (i) an ordered magnetic field with strength 
of the order of $ 10^{14}-10^{15}$ G must be present at the flow injection point and (ii) 
the ratio of the neutrino-driven mass flux to magnetic flux must be sufficiently low. 
The possible existence of such strong magnetic fields in systems like those mentioned above
appears to be indicated by observations, and is also supported by recent numerical simulations.
However, the dependence of the neutrino-driven mass loading
on the conditions in the disk and on the geometry of the  magnetic field is yet
an open issue.

A common approach in studies of magnetized disk outflows is to seek self-similar
solutions of the trans-field equation (e.g.,  Refs.~\cite{Li92,Contopoulos94,Vlahakis03}).  
While this approach allows the magnetic field geometry 
to be calculated self-consistently, incorporation of gravity (in the relativistic case) 
as well as neutrino heating in the flow equations is precluded.
Furthermore, it does not allow simple matching of the self-similar outflow solution
to a Keplerian disk (but c.f., Ref.~\cite{Vlahakis03}).  The immediate implication is 
that  the mass-to-magnetic flux ratio, which is determined by the regularity condition
at the slow magnetosonic point, remains a free parameter.  One must therefore adopt a different 
approach.   One method is to seek solutions of the flow equations that pass smoothly
through the slow magnetosonic point, assuming the magnetic field geometry is given.
Such an analysis has been carried out recently for winds from a disk around a 
hyper-accreting black hole \cite{Levinson06}. A brief account of the model and results 
is given below. 

\section{The Model}
We consider a stationary, axisymmetric MHD 
wind expelled from the surface of a hot, magnetized disk surrounding a 
non-rotating black hole.  The range of conditions in the disk is envisioned to be
similar to that computed by Popham et. al. \cite{Popham99} for hyperaccreting black holes
with mass accretion rates $10^{-1} - 10$  $M_{\odot}$ s$^{-1}$ and  viscosity parameters 
$\alpha_{vis}=0.01 - 0.1$.   Under such conditions the dominant cooling mechanism in the
disk is neutrino emission.  The major fraction of the neutrino luminosity is generated 
in the inner disk regions, within 10 Schwarzschild radii or so, and
lies in the range $L_\nu=10^{51}$ - $10^{54}$ erg s$^{-1}$ for the above range of
accretion rates and viscosity parameter.

The model calculates the structure of the GRMHD outflow below the slow magnetosonic point
for a given magnetic field geometry,
treating the neutrinos emitted from the disk as an external energy and momentum source.
To simplify the analysis the neutrino source is taken to be spherical with radius $R_\nu$.
The model is characterized by three parameters: the black hole mass $M_{BH}$,
the neutrino luminosity $L_{\nu}$, and the neutrinospheric radius $R_{\nu}$.
It solves a set of coupled ODEs that are derived from the general 
relativistic  energy-momentum equations, describing the change along a 
given streamline $\Psi(r,\theta)=$ const of the specific energy ${\cal E}$ , 
specific entropy $s$, and poloidal flow velocity $u_p$:
\begin{eqnarray}
(\rho/m_N)k_BT s^\prime=-u_\alpha q^\alpha,\label{s}\\
\rho c^2 {\cal E}^\prime=-q_t,\label{E=q}\\
(\ln u_p)^\prime={F\over D}\label{u_p}.
\end{eqnarray}
Here  $^\prime$ denotes derivative along streamlines $u^\alpha\partial_\alpha$,
$q^{\beta}$ denotes the source terms associated with energy and momentum exchange
with the external neutrino source, $u^\alpha$ is the outflow 4-velocity, $\rho$ is the 
baryon rest mas  density and $T$ is the temperature.   The denominator on the R.H.S.
of Eq. (\ref{u_p}) is given explicitly as $D=-(\alpha^2-R^2\Omega^2-M^2)^2(u_p^2-u_{SM}^2)
(u_p^2-u_{FM}^2)/u_A^2$,
where $u_A$, $u_{SM}$ and $u_{FM}$ are the Alfv\`en, slow and fast magnetosonic wave speeds, 
respectively, $\alpha$ is the lapse function, $\Omega$ is the angular velocity defined below, 
$R$ is the cylindrical radius and $M=u_p/u_A$ is the Alfv\`en Mach number.
The term $F$ can be expressed as
$F=\zeta_1(\ln B_p)^\prime+\zeta_2(\ln \alpha)^\prime+\zeta_3(\ln R)^\prime+
\zeta_4(\ln {\cal E})^\prime+\zeta_5 (\ln s)^\prime$, where the coefficients $\zeta_k$
are functions of the flow parameters, given explicitly in Ref.~\cite{Levinson06}, 
and $B_p$ is the poloidal field component.  Since the derivatives 
$(\ln B_p)^\prime$ and $(\ln R)^\prime$ depend on the magnetic field geometry
which is unknown a priori, additional equation is needed.  Our approach is
to invoke a given field geometry.  To examine the dependence of mass flux on the latter,
we obtained solutions for different magnetic field configurations,
focusing particularly on split monopole and $r$ self-similar geometries.
Equations (\ref{s})-(\ref{u_p}) are augmented by an equation of state for the mixed fluid of baryons,
photons and electron-positron pairs.  In addition there are three integrals of motion
of the MHD system: the mass-to-magnetic flux ratio $\eta(\Psi)$, the angular velocity of magnetic 
field lines $\Omega(\Psi)$, and the specific angular momentum ${\cal L}(\Psi)$.
The two invariants $\Omega(\Psi)$ and ${\cal L}(\Psi)$ are fixed by a choice of 
boundary conditions.  The mass flux $\eta(\Psi)$ is an eigenvalue of the 
problem.  The integration starts sufficiently deep in the disk atmosphere where the 
density is high, the flow velocity is very small, and the ratio of baryonic pressure
and the light fluid pressure is about unity so that the specific entropy
is dominated by the baryons.
The value of $\eta$ is adjusted iteratively by repeating the integration, changing the 
boundary value of the poloidal velocity $u_{p0}$ each time, until a smooth transition across 
the slow magnetosonic point is obtained.

\section{Results}
Typical solutions are shown in Figs 1 and 2.  Those were obtained using
a split monopole geometry (see Ref.~\cite{Levinson06} for more details).
In Fig. 1 the change along a field line of the flow quantities indicated is plotted 
against the normalized height above the disk midplane, $z/R_0$, where
$R_0$ is the radius at which the field line meets the disk.
The inclination angle of the field line in this example is about $12^\circ$
with respect to the rotation axis, so it lies in the so-called {\it stable} regime.
The mass flux is thermally driven in this regime rather than centrifugally driven.
As seen, the slow magnetosonic point is located at $z\simeq R_0$, which we find
to be typical for injection along low inclination field lines.  
The integration started well beneath the slow magnetosonic point where
the specific entropy is dominated by the baryons.  In this region
adiabatic cooling is negligible and kinetic equilibrium is established. The net energy 
deposition rate nearly vanishes, viz., $q^t(T_0)\simeq0$, and the temperature
is $T_0\simeq T_\nu$, where $T_\nu$ is the neutrino equilibrium temperature.
As the flow accelerates its 
temperature starts falling, and since the neutrino 
cooling rate is very sensitive to the temperature $q^t$ increases rapidly leading to the 
steep rise of the entropy per baryon seen in Fig 1.  The terminal value of $s$ is however
modest, $s/k_B<100$ in all cases examined, so the net outflow energy is practically not 
affected.  The main effect of neutrino heating is not to change the 
net outflow energy along streamlines, but rather to enhance the baryon load by 
increasing the light fluid pressure in layers of higher baryon density.   
Thus, the initial value of ${\cal E}$ presents, essentially, an upper limit
for the asymptotic Lorentz factor of the wind. 
The values of the mass flux, which we somewhat arbitrarily define as 
$\dot{M}=\rho_0u_{p0}2\pi R_0^2$ (for a two-sided
outflow), that corresponds to the cases depicted in Fig. 1 are
$10^{28}$ and $7\times10^{30}$ g s$^{-1}$ for $T_\nu=2$ and 4 MeV, respectively, and the
the corresponding values of the total energy per baryon are
${\cal E}=5\times10^2m_pc^2$ and $2m_pc^2$ for a surface poloidal field strength of $10^{15}$ Gauss.  

The dependence of the mass flux on the inclination angle of the field line is shown in Fig. 2.
As seen, the mass flux is quite sensitive to the inclination of the field line.
 For field lines in the {\it unstable} 
regime (inclination angles larger than $30^\circ$ to the vertical) the wind is centrifugally driven
and, for the parameter range explored, is found to be always subrelativistic, owing to
the large neutrino-driven mass flux expelled along the field lines.  The slow magnetosonic
point in this case is located very near the surface, at $z/R_0<<1$.  A systematic study of the 
dependence of $\eta$ on the various parameters is presented in Ref.~\cite{Levinson06}.

\begin{figure}
  \includegraphics[height=.3\textheight]{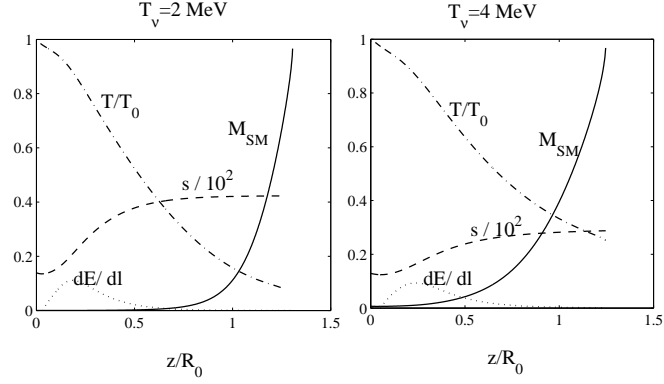}
  \caption{Profiles of various quantities in the 
sub-slow magnetosonic region, computed using a split monopole magnetic field
with $\tan\theta=0.2$, where $\theta$ is the inclination angle of the field line 
with respect to the symmetry axis, Keplerian rotation $\Omega=\Omega_k$, and 
surface magnetic field $B_{p0}=-10B_{\phi0}=10^{15}$ G.  Each panel corresponds to a run with a 
different neutrino luminosity (indicated in terms of the effective 
black-body temperature $T_\nu$).  
The quantities plotted in each panel are: the slow magnetosonic Mach number $M_{SM}$ (solid line),
the dimensionless entropy per baryon $s$ (dashed line), the temperature $T$ in units of the initial 
temperature $T_0$ (dotted-dashed line), and the energy deposition per baryon per unit 
length along the streamline measured in units of $m_pc^2/R_0$, $d{\cal E}/dl$ (dotted line).
All quantities are given as functions of the normalized height above the disk midplane $z/R_0$.}
\end{figure}

\begin{figure}
  \includegraphics[height=.3\textheight]{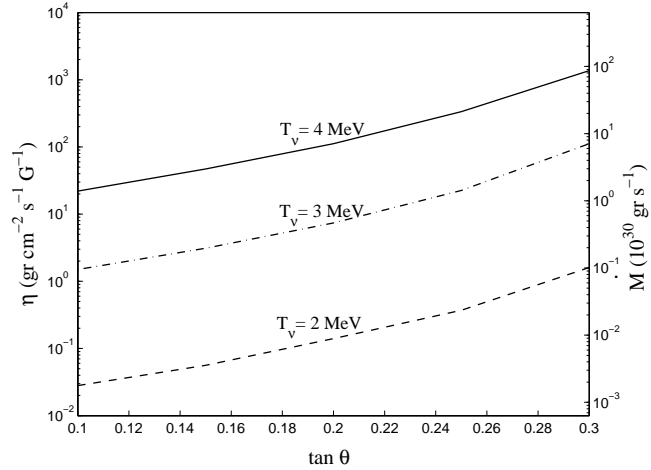}
  \caption{Dependence of mass-to-magnetic flux ratio $\eta$ (left axis) on
the inclination angle $\theta$ of the split-monopole field line.
The corresponding values of the mass flux $\dot{M}$ are indicated on the right axis}
\end{figure}

\section{Conclusion}

Our main conclusion is that ejection of relativistic outflows from the 
innermost disk radii, within several $r_s$ or so, is possible in principle for certain magnetic field 
configurations even in non-rotating black holes, provided the neutrino luminosity 
emitted by the disk does not exceed $10^{52}$ ergs s$^{-1}$ or so, and the magnetic 
field is sufficiently strong ($B_p\sim10^{15}$ G) .  The neutrino-driven mass flux depends 
rather sensitively on the neutrino luminosity, on magnetic field geometry in 
the sub-slow magnetosonic region, and on the angular velocity of magnetic flux
surfaces, but is highly insensitive to the strength of poloidal and toroidal 
magnetic field components near the surface provided the Alfv\'en Mach number 
is sufficiently small there.  The picture often envisioned, 
of an ultra-relativistic core surrounded by a slower, baryon rich wind 
\cite{Levinson93,Levinson00} seems a natural consequence of a neutrino-assisted 
MHD disk outflow.

The hyperaccretion process is likely to be intermittent, leading to temporal changes
in the neutrino luminosity.  This should result in large variations of the Lorentz
factors of consecutive fluid shells expelled from the disk in the polar region, owing to 
the sensitive dependence of mass loading on $L_\nu$.  In this situation we expect strong shocks
to form in the outflow.  If the polar disk outflow is associated with the GRB-producing 
jet, then the observed gamma-ray emission can be quite efficiently produced behind those shocks.
   
Finally, the conditions we find to be optimal for launching an ultra-relativistic jet
in the polar region, are also the conditions favorable for large neutron-to-proton 
ratio in the disk.  However, for the steady flow considered above 
we estimate the ratio of the neutronization timescale and flow time to
be of order unity, and so the electron fraction is expected to evolve as the flow accelerates.
Detailed analysis of the composition profile in the outflow is left for future work.

\begin{theacknowledgments}
I thank Arieh Konigle for inspiring conversations.
This work was supported by an ISF grant for the Israeli Center for High Energy Astrophysics.
\end{theacknowledgments}

\end{document}